\documentclass[journal]{IEEEtran}
\usepackage{amsmath,amssymb,amsfonts}
\usepackage{algorithmic}
\usepackage{algorithm}
\usepackage{array}
\usepackage[caption=false,font=normalsize,labelfont=sf,textfont=sf]{subfig}
\usepackage{textcomp}
\usepackage{stfloats}
\usepackage{url}
\usepackage{verbatim}
\usepackage{graphicx}
\usepackage{cite}
\usepackage{adjustbox}
\usepackage{multirow}
\usepackage[table]{xcolor}
\usepackage{booktabs}

\begin{document}

\title{Optimal Energy Management in Indoor Farming Using Lighting Flexibility and Intelligent Model Predictive Control}
\author{Mohammadjavad Abbaspour$^1$, Mukund R. Shukla$^2$, Praveen K. Saxena$^2$, and Shivam Saxena$^1$ \\
	$^1$Department of Electrical and Computer Engineering, University of New Brunswick, Canada \\
	$^2$Department of Plant Agriculture, University of Guelph, Canada}
\maketitle
\begin{abstract}
Indoor farming enables year-round food production but its reliance on artificial lighting significantly increases energy consumption, peak load charges, and energy costs for growers. Recent studies indicate that plants are able to tolerate interruptions in light, enabling the design of 24-hour lighting schedules (or "recipes") with strategic light modulation in alignment with day-ahead pricing. Thus, we propose an optimal lighting control strategy for indoor farming that modulates light intensity and photoperiod to reduce energy costs. The control strategy is implemented within a model predictive control framework and augmented with transformer-based neural networks to forecast 24-hour ahead solar radiation and electricity prices to improve energy cost reduction. The control strategy is informed by real-world experimentation on lettuce crops to discover minimum light exposure and appropriate dark-light intervals, which are mathematically formulated as constraints to maintain plant health. Simulations for a one-hectare greenhouse, based on real electricity market data from Ontario, demonstrate an annual cost reduction of \$318,400 (20.9\%), a peak load decrease of 1.6 MW (33.32\%), and total energy savings of 1890 MWh (20.2\%) against a baseline recipe. These findings highlight the potential of intelligent lighting control to improve the sustainability and economic feasibility of indoor farming.
\end{abstract}


\section{Introduction}
\label{sec:introduction}
Indoor farming in climate-controlled greenhouse environments is gaining traction as it mitigates the impact of volatile weather, however, it consumes approximately 15–20 times more energy than traditional outdoor farming methods \cite{horomia2021global}. Major sources of energy consumption are heating, ventilation, and air conditioning (HVAC) and supplementary artificial lighting, which account for over 80\% of the demand  \cite{horomia2021global, osman2024synergy, christensen2020agent}. While these systems enable growers to operate year-round, the resultant substantial energy costs negatively impact food prices and raise serious concerns regarding the long-term sustainability of indoor farming.


To reduce energy costs in controlled environment agriculture, past studies have looked at shifting electricity use, lowering peak demand charges, and using energy more efficiently during price changes. A common method is to use distributed energy resources (DERs) like solar panels and batteries \cite{achour2020supervisory, fu2024agri, zhuang2018stochastic}. These studies show that DERs can help lower energy costs, but the high cost of solar and batteries is still a major barrier, especially for small, and medium-scale growers who have already spent a lot on artificial lighting \cite{mueller2018unlocking}.


On the other hand, plant physiology research suggests that plants can tolerate intermittent lighting if their total daily light requirement is met. This allows dynamic lighting control strategies that adjust energy use based on electricity prices and peak demand \cite{mohagheghi2023energy, alaviani2022optimal}, thus reducing costs without harming plant health. Studies mainly focus on two strategies: (1) using intermittent lighting while maintaining total daily light intake, and (2) lowering daily intake to reduce energy use while preserving yield. As noted in our previous work \cite{abbaspour2024new}, the first approach shifts energy consumption based on time-varying prices, while the second approach directly reduces energy consumption. These approaches result in lighting schedules, or lighting "recipes" that define specific light intensities at defined time intervals over 24 hours, thus allowing precise and flexible control of artificial lighting in alignment with day-ahead energy pricing.

Within the first approach of maintaining consistent daily light intake, studies  indicate that longer intervals can benefit plant growth, as long as sufficient lighting is maintained without prolonged exposure to high light intensity \cite{elkins2020longer, shao2020effects}. This method also enables electricity cost management by adjusting light intensity, light/dark cycle durations, and the distribution of light/dark intervals in response to peak and off-peak electricity pricing. The second approach, involving lowering the daily light intake, depends on factors such as plant species, growth stage, and environmental conditions. Several studies have examined its effectiveness. For example, Schwend \textit{et al.} proposed a dynamic lighting schedule with a reduced light intensity setpoint using rule-based control, resulting in a 21.1\% reduction in artificial lighting energy use without affecting plant fresh weight \cite{schwend2016test}. Lork \textit{et al.} designed an optimization algorithm to minimize energy costs for LED lighting in lettuce production, achieving cost reductions of 40–52\% \cite{lork2020minimizing}, while studies on other plant species have shown that moderate lighting conditions can enhance energy efficiency without compromising plant health \cite{avgoustaki2020basil, avgoustaki2021energy, an2021evaluation}.

While earlier contributions have significantly advanced lighting flexibility strategies to improve energy management, they predominantly rely on rule-based methodologies that cannot guarantee globally optimal outcomes \cite{schwend2016test, avgoustaki2020basil}. Few studies have employed optimization-based methods to design lighting strategies \cite{lork2020minimizing, alaviani2022optimal, mohagheghi2023energy, avgoustaki2021energy}, and while these approaches effectively reduce energy costs associated with artificial lighting, they exhibit three major limitations. First, existing studies do not account for other greenhouse loads, resulting in a suboptimal minimization of fundamental cost components such as peak demand charges and electrical infrastructure usage. Where some papers integrate multiple loads, they do not fully leverage the flexibility potential of lighting control \cite{zhuang2018stochastic, fu2024agri, achour2020supervisory, bozchalui2014optimal, avgoustaki2021energy}. Second, current optimization strategies for 24-hour-ahead lighting schedules often rely on rough hourly or daily predictions of solar radiation and electricity prices. These coarse estimates overlook short-term fluctuations, leading to suboptimal schedules that miss opportunities for cost savings and efficient energy use. Third, these studies neither evaluate the impact of the designed lighting recipes on plant health nor incorporate key physiological parameters, such as minimum required light intake, suitable light/dark cycle durations, and balanced light/dark distribution, into their optimization frameworks. As a result, the proposed strategies may fail to ensure optimal conditions for plant growth.

To address these gaps, we propose a novel approach to generating day-ahead optimized lighting schedules using electricity price and solar radiation predictions. Our method aims to reduce energy costs associated with both hourly energy consumption and peak demand charges while ensuring minimum plant health requirements, including daily light intake and adequate light/dark intervals. We conduct real-world experiments on lettuce crops to establish a minimum threshold for daily light intake and formulate the resulting plant physiology constraints as an optimization problem, solved using model predictive control (MPC). Additionally, we enhance the MPC with a novel transformer-based predictor for electricity prices and solar radiation, improving the lighting schedule’s efficiency and maximizing cost savings while reducing peak demand. Furthermore, we simulate other greenhouse loads to assess their impact on peak demand reduction, enabling a more comprehensive evaluation of energy cost and consumption in a simulated greenhouse with adjustable parameters. Accordingly, our contributions are fourfold:
\begin{itemize}
	\item Establishing new boundaries for minimum required daily light intake and allowable light intensity via real-world experimentation to enable the design of flexible lighting recipes.
        \item Formulating the boundaries as mathematical constraints within a MPC framework to generate day-ahead recipes that reduce energy costs while maintaining plant health.
        \item Extending MPC by designing novel transformer-based electricity price and solar radiation predictions to maximize energy savings
	\item Integrating greenhouse loads within the MPC objective function to optimize peak demand reduction and analyze energy cost and consumption under adjustable conditions.
\end{itemize}
The organization of the paper is as follows: Section II provides background on the interaction between various lighting strategies and their impact on plant growth. Section III outlines the system modeling, including the mathematical formulation of the MPC, modeling of greenhouse loads, and the experimental methodology. It also covers the design of lighting recipes and the definition of constraints for the MPC. Section IV presents and discusses the results, while Section V concludes the paper and proposes directions for future research.
\section{Background}
\label{sec:Background}
\begin{figure}
	\centering
	\scalebox{0.12}{\includegraphics{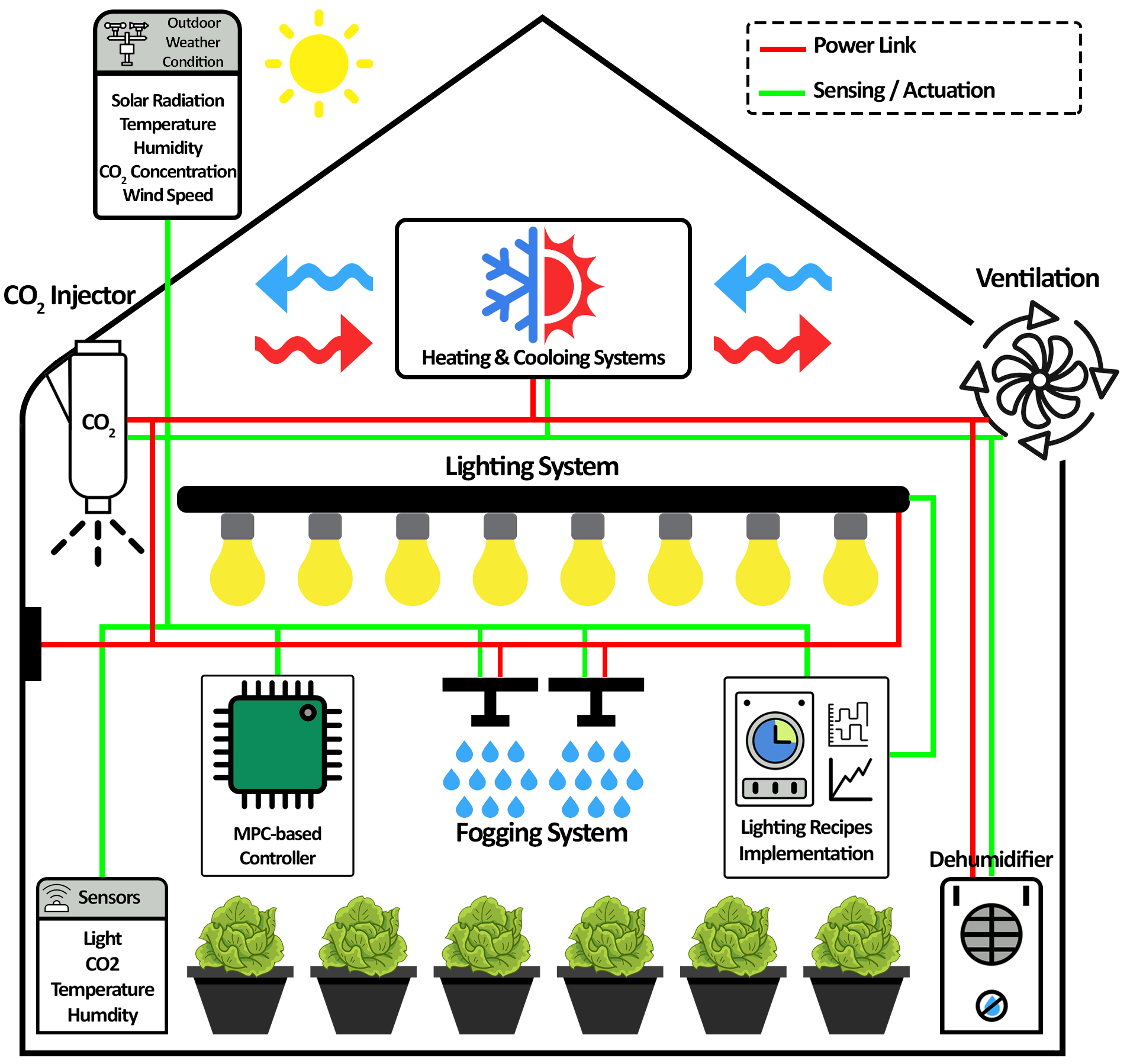}}
	\caption{Schematic representation of a greenhouse.}
	\label{fig1}
\end{figure}

This section provides background on how plant physiological factors influence energy management strategies, with a focus on optimizing lighting in greenhouse environments. Fig. \ref{fig1} illustrates the key processes, inputs, and outputs of the greenhouse system.

\subsection{Plant Physiology}
\label{sec:PlantPhysiology}
Plants utilize light as a primary energy source for photosynthesis, a process that converts light energy into chemical energy necessary for growth and development. Photoreceptors within plant cells detect specific wavelengths, triggering physiological and developmental responses that optimize adaptation to environmental conditions. However, environmental stress can disrupt these processes, impairing photosynthesis and overall plant health \cite{jones2018using}. It should be noted that plants require both light and dark periods for optimal functioning.

\begin{figure}
	\centering
	\scalebox{0.28}{\includegraphics{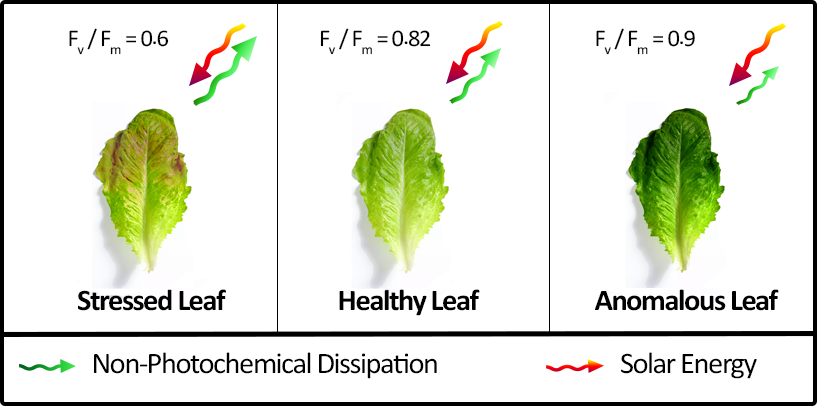}}
	\caption{Fluorescence response of three leaves under different conditions.}
	\label{fig2}
\end{figure}

Photosynthesis is driven by the pigment chlorophyll, which absorbs light and re-emits some of the absorbed energy as fluorescence, which is a useful indicator for photosynthetic efficiency and plant stress.  Key fluorescence parameters include $F_o$, representing the baseline fluorescence emission in dark-adapted conditions, and $F_m$, which indicates the maximum fluorescence emission when photosynthetic processes are reduced due to excess energy dissipation. Additionally, variable fluorescence (\(F_v\)), calculated as the difference between \(F_m\) and \(F_o\), along with the ratio \(F_v/F_m\), provides further insights into plant stress responses and the overall functionality of the photosynthetic apparatus. A decline in these values often signifies stress-induced damage. Using these measurements, we can assess the tolerance of plants when light intensity is reduced. Fig. \ref{fig2} illustrates how excess non-photochemical dissipation (indicated by high $F_m$) can cause stress-related consequences, such as tipburn, while very low $F_m$ values may lead to abnormal conditions for plant growth. For most plants, the normal range for the $F_v/F_m$ ratio is between 0.79 and 0.85 \cite{romermann2016plant, bi2024investigating}.

\subsection{Controlled Environment Agriculture System}
\label{sec:ControlledEnvironment}
A Controlled Environment Agriculture System (CEAS) integrates essential environmental control mechanisms to regulate light, temperature, $\text{CO}_2$ concentration, and humidity, while also accounting for external factors such as solar radiation and wind speed. It relies on a network of interconnected subsystems that represent the most common energy-consuming loads in climate-controlled greenhouses, as described below.
\subsubsection{Temperature Control}
The greenhouse temperature is regulated using a combination of the heater, chiller, and ventilator. The temperature balance equation is given by:
\begin{flalign}
	&&C_{\text{air}} \frac{dT_{\text{in}}}{dt} = Q_{\text{solar}} + Q_{\text{heater}} - Q_{\text{chil}} - Q_{\text{vent}} -Q_{\text{conv}}
\end{flalign}
where $C_{\text{air}}$ represents the heat capacity of indoor air, $Q_{\text{solar}}$ corresponds to the heat gain from solar radiation, $Q_{\text{heater}}$ and $Q_{\text{chil}}$ denote the thermal contributions of the heating and cooling systems, respectively, $Q_{\text{vent}}$ accounts for heat loss due to air exchange with the external environment, and $Q_{\text{conv}}$ represents convective and conductive heat loss through the greenhouse structure.

Many studies have provided the formulations for $Q_{\text{ventilation}}$, $Q_{\text{convection}}$, and $Q_{\text{solar}}$, as referenced in \cite{bozchalui2014optimal, achour2020supervisory}. Additionally, $Q_{\text{heater}}$ or $Q_{\text{chiller}}$  can be computed and adjusted to maintain the temperature within the desired range. 
\subsubsection{Humidity Control}
Humidity control promotes plant growth and prevents fungal disease via fogging systems, dehumidifiers, and ventilators. The humidity balance equation is:
\begin{flalign}
	RH_{\text{in}}(t) = \frac{P_{\text{bar}}(t)}{P_{\text{sat}}(t)} 100 \%
\end{flalign}
where $RH_{\text{in}}$ is the relative humidity at time $t$, $P_{\text{bar}}$ is the actual vapor pressure, and $P_{\text{sat}}$ is the saturated vapor pressure. The relative humidity varies over time as a function of the moisture in the air, represented by the balance equation below. 
\begin{flalign}
	\frac{dRH_{\text{in}}}{dt} = G_{\text{fog}} + G_{\text{dehum}} - G_{\text{ven}}
\end{flalign}
where $G_{\text{fog}}$ represents the moisture added by the fogging system, $G_{\text{dehum}}$ is the moisture removed by the dehumidifier, and $G_{\text{vent}}$ accounts for the humidity exchange between the greenhouse and the outside air.
\subsubsection{$\text{CO}_2$ Control}
$\text{CO}_2$ is an essential nutrient for plants and crucial for photosynthesis. While atmospheric $\text{CO}_2$ concentration is typically around 380 ppm, elevated $\text{CO}_2$ levels in advanced greenhouses can significantly enhance photosynthesis. The variation in indoor $\text{CO}_2$ levels is described below.
\begin{flalign}
	\frac{d\text{CO}_2}{dt} = J_{\text{inj}} - J_{\text{vent}}
\end{flalign}
where $J_{\text{inj}}$ is the rate of $\text{CO}_2$ injection and $J_{\text{vent}}$ represents $\text{CO}_2$ loss due to air exchange.
\subsubsection{Lighting Control}
    Indoor lighting can account for 30–80\% of a greenhouse’s total energy consumption \cite{osman2024synergy, christensen2020agent}. A key parameter in greenhouse lighting management is a plant’s daily light integral (DLI), which quantifies the total amount of light received in a day and is expressed in \(\frac{\text{mol}}{{m^2} \cdot \text{day}}\). DLI is determined by the photosynthetic photon flux density (PPFD), measured in \(\frac{\mu \text{mol}}{{m^2}s}\), and the total daily light duration (TDLD), which refers to the number of light hours per 24-hour cycle. In simplified terms, PPFD is analogous to instantaneous power, the photoperiod represents the duration of light exposure, and DLI is the cumulative energy delivered over time. A lighting recipe, denoted as $y(t)$, can be mathematically modeled by integrating PPFD and photoperiod over a discretized 24-hour period, as follows:
\begin{flalign}
	& y(t) = \sum_{n=1}^{N_I} r[n] (u(t-(n-1)I)- u(t-nI)) \\
	& T = I \;\sum_{n=1}^{N_I} u(r[n]) \\
	& D = \int_{t=0}^{N_I \times I} 3600 \times 10^{-6} \quad y(t) \quad dt
\end{flalign}
where $r[n]$ is the PPFD value at time interval $n$, $I$ is the length of each of the $N_I$ identical time intervals that make up a 24-hour period, and $T$ and $D$ represent the TDLD and DLI, respectively, while $u(t)$ denotes the unit step function. Among the environmental parameters, the flexibility offered by lighting system is the highest and allows for the design of different recipes to reduce energy costs.
\subsubsection{Electricity Costs of Greenhouse Loads}
\label{sec:ElectricityConsumptionandCosts}
To calculate the electricity consumption of CEAS subsystems introduced in Section \ref{sec:ControlledEnvironment}, short time intervals, $I$, must be defined as in Equations (5)-(7). During each interval, controllers regulate light, temperature, relative humidity, and $\text{CO}_2$ levels. They determine the number of devices that are turned on and the setpoint power of each device. Using this information, Equation (5) can be applied to compute electricity consumption per interval.
\begin{flalign}
	&E_{I} = \sum_{l=1}^{M} \sum_{n=1}^{n^l} w^{l,n} \; P_{\text{rated}}^{l} \; I 
\end{flalign}
where, $M$ represents the total number of CEAS subsystems, while $E_I$ denotes the total energy consumption. The term $P_{\text{rated}}^l$ corresponds to the maximum power rating of each device, $n^l$ specifies the number of devices per load, and $w^{l,n}$ serves as the scaling factor for each device. 

Assuming that the lighting system maintains a consistent efficiency in converting electricity to light within its operational PPFD range, its power consumption can be approximated as a linear function of PPFD. Based on this assumption of linearity, Equation (8) is derived from Equations (5)–(7). Consequently, the power consumption of a given lighting recipe at time $t$, expressed as \(P_n(y(t)) = w^{1,n} \cdot P_{\text{rated}}^{1}\)(where $l=1$ corresponds to the lighting system), is directly proportional to the artificial light component $r_a[n]$, as given by the equation \(r[n] = r_a[n] + r_s[n],\) where $r_s[n]$ represents the natural light contribution.  

A greenhouse's total electricity cost consists of energy price/tariff, peak demand (PD), and infrastructure and cost recovery adjustment (ICRA). The monthly cost is determined by:  
\begin{flalign}
&\hspace{-1em} C_m = \sum_{n=1}^{N_I} (P_{ep,n} + P_{icra}) \cdot E_{I,n} + P_{pd} \max_{1\leq n \leq N_I} {E_{I,n}}
\end{flalign}
where $C_m$ denotes the total monthly cost over $N_I$ intervals, $E_{I,n}$ represents the energy consumed per interval, $P_{ep,n}$ is the electricity price, and $P_{icra}$ is the monthly ICRA per kWh. All monetary values are expressed in \$/kWh and kWh.
\section{System Design}
\label{system_Modeling}
\begin{figure}
	\scalebox{0.23}{\includegraphics{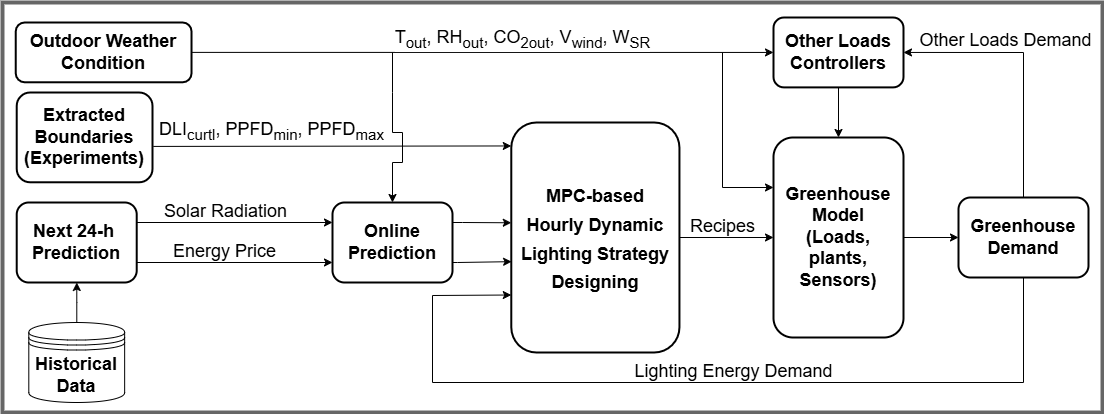}}
	\caption{High-level block diagram of the system modules and their interactions.}
	\label{fig3}
\end{figure}

This section introduces the design of the proposed system, including the methodology for designing trial recipes to identify plant physiology boundaries, the modules to predict electricity price and solar radiation, as well the integration of the aforementioned components within an MPC-based optimization framework to generate daily recipe to reduce energy costs. Fig. \ref{fig3} summarizes the system’s key modules and their inputs/outputs.

\subsection{Experimental Methodology and Design of Recipes}
\label{Experiments}
A key objective in establishing the flexibility of plants to tolerate light interruptions is to discover boundaries for the minimum DLI, TDLD, and appropriate light/dark intervals. We thus developed a general experimental methodology to find these values through trial recipe generation and execution across plant growth chambers that regulated temperature, ventilation, and humidity, while we isolated DLI and TDLD as control variables. Our methodology elected to maintain a constant DLI across all chambers while adjusting TDLD to reduce the number of physical trials, since modifying TDLD indirectly affects both PPFD and light/dark interval duration. It is worth noting that our methodology intentionally sets the DLI to a lower value ($\text{DLI}_{\text{Curtl}}$) than conventionally noted in literature to test the impact of stress on the plant and resultant potential energy savings potential.

Our methodology then designs recipes by fine-tuning intensity, photoperiod, and light/dark intervals while measuring the impact of the variation on plant stress, measured by observations of $F_v/F_m$. To this end, the PPFD is constrained within the range $[\text{PPFD}_{\text{min}}, \text{PPFD}_{\text{max}}]$ to sustain photosynthetic function while avoiding light-induced stress. The photoperiod is subsequently modified to meet the required $\text{DLI}_{\text{Curtl}}$. Additionally, constraints on the minimum and maximum consecutive light intervals, denoted as $I_{L,\text{min}}$ and $I_{L,\text{max}}$, along with the minimum and maximum consecutive dark intervals, $I_{D,\text{min}}$ and $I_{D,\text{max}}$, are applied to maintain biologically appropriate light-dark cycles essential for plant development. Note that, to respect physiological limits, the minimum interval length is defined as $I = max(I_{L,\text{min}}, I_{D,\text{min}}) $. Accordingly, we formulate the following mathematical constraints to be foundational components in the design of a recipe:
\begin{flalign}
& \text{PPFD}_{\text{min}} \leq y(t) \leq  \text{PPFD}_{\text{max}} \quad \forall \quad t \in [0, N_I\times I] \\
& D = DLI_{\text{Curtl}}  \\
& \delta(r[n]) =  0  \quad \text{if} \quad \sum_{k = n - \left(\frac{I_{L,\text{max}}}{I}\right)}^{n - 1} \delta(r[k]) > \frac{I_{L,\text{max}}}{I} \\
& \delta(r[n]) =  1  \quad \text{if} \quad \sum_{k = n - \left(\frac{I_{D,\text{max}}}{I}\right)}^{n - 1} \delta(r[k]) > \frac{I_{D,\text{max}}}{I}
\end{flalign}
\subsection{Transformer-Based Prediction Models for Electricity Price and Solar Radiation}
\label{Predicions}
The design of lighting recipes require accurate predictions of solar radiation and electricity prices, where we propose the use of transformer-based neural networks that leverage historical data for time-series forecasting. Transformers are well-suited for time series forecasting due to their ability to model long-range dependencies, capture temporal patterns, handle multiple correlated time series (e.g., weather, demand, and pricing), and perform accurate multi-horizon forecasting \cite{li2019enhancing}. 


The electricity price prediction model incorporates hourly features such as day of the week, hour of the day, year, temperature, wind speed, season, market demand, public holiday indicator, and energy generation from nuclear, gas, hydro, wind, solar, and biofuel sources. The solar radiation prediction model utilizes similar time-based variables along with meteorological parameters, including temperature, dew point, wind speed, station pressure, sea-level pressure, wind direction, and relative humidity. Both models predict the next 24 hours based on the preceding 24-hour data window, enabling real-time adjustments to improve algorithm performance in reducing cost and energy usage.  

Each model employs a transformer architecture suited for time-series forecasting, with multiple layers, attention heads, and feedforward dimensions. The solar radiation model consists of 3 transformer layers, while the electricity price model uses 4 layers, both with 4 attention heads, a model dimension of 64, a feedforward dimension of 256, a dropout rate of 0.1, and ReLU activation. Layer normalization and residual connections are integrated for stability.  

To enhance data quality and minimize the impact of extreme values, a quantile-based approach was applied for outlier removal. The interquartile range (IQR), which quantifies the spread of data by measuring the difference between the first quartile ($Q_1$) and the third quartile ($Q_3$), was used to identify and remove outliers. We set $Q_1$ and $Q_3$ at the $10^{\text{th}}$ and $80^{\text{th}}$ percentiles of the data distribution, respectively. The $IQR$ was then computed as:  
\begin{flalign}
	&IQR = Q_3 - Q_1 \nonumber \\
	&LB = Q_1 - k \times IQR \quad , \quad  UB = Q_3 + k \times IQR 
\end{flalign}
where $k$ is set to 1.5. Data points beyond these bounds were excluded to manage extreme variations while preserving essential fluctuations.  

A sliding window approach with a 24-hour sequence length was applied for training, with the Adam optimizer and weight decay. Mean Squared Error (MSE) was used as the loss function, while Root Mean Squared Error (RMSE) and Mean Absolute Error (MAE) assessed performance. The models were trained on a nine-year dataset, using a 70-20-10 train-validation-test split.  
\subsection{Generation of Optimal Lighting Recipes}
The generation of an optimal lighting recipe is mathematically formulated as an optimization problem aimed at minimizing energy costs, subject to the physiological plant constraints identified in the previous subsection (10)-(13). We adopt an MPC approach to iteratively determine the artificial lighting recipe over a full day ($N = 24$), forecasting the remaining time steps at each iteration. Equations (15)–(25) define the components of the optimization at step $i$. Equation (15) represents the main objective function, which consists of three terms:
\begin{flalign}
\hspace{-2em}\min_{\substack{\{v[n], w[n]\}_{n=1}^{N_I} \\ \{u[n], x[n]\}_{n=i}^{N_I}}}  
& \quad \alpha \sum_{n=1}^{N_I} P_{ep}[n] r_a[n]  
+ \beta \sum_{n=1}^{N_I} r_a^2[n]  \nonumber \\ 
&+ \gamma \max_{1\leq n \leq N_I} \{r_a[n]\} 
\end{flalign}
where, $P_{ep}[n]$ represents the integration of real values for $n < i$ and predicted values for $n \geq i$ for electricity price. The artificial lighting component of the lighting recipe, $r_a[n]$, consists of a binary variable $u[n]$, which indicates the presence or absence of artificial lighting and is determined by the optimizer. Additionally, $x[n]$ represents the intensity of artificial lighting, incorporating real values for $n < i$ and predicted values for $n \geq i$. The hyperparameters $\alpha$, $\beta$, and $\gamma$ define the importance of each term. Accordingly, the first term in the objective function minimizes the electricity cost associated with artificial lighting. The second term ensures that the resulting lighting recipe is smoother and more suitable for plant growth. The third term regulates peak demand, thereby reducing the peak demand in months where artificial lighting is responsible for the monthly peak. Note that the optimization problem is both nonlinear and convex, allowing it to be solved using convex optimization techniques, such as interior-point or gradient-based methods.

We constrain the PPFD range, $[\text{PPFD}_{\text{min}}, \text{PPFD}_{\text{max}}]$, to ensure effective photosynthesis and avoid stress from excessive light. Within these bounds, photoperiods, optimized during scheduling, are defined in one-hour increments with a minimum duration of $I = 1$ hour. Equations (16)–(25) represent the corresponding constraints.

\begin{flalign}
& r_a[n] = x[n]\; u[n], \quad n = 1, \dots, N_I \\
& u[n] = u_{\text{opt}}[n], \quad x[n] = x_{\text{opt}}[n], \quad n = 1, \dots, i-1 \\
& u[n] \in \{0,1\}, \quad x[n] \geq \text{PPFD}_{\text{min}}, \quad  n = i, \dots, N_I \\
& r_s[n] = Y[n]\; v[n]\; w[n], \quad n = 1, \dots, N_I \\
& v[n] \in \{0,1\}, \quad 0\leq w[n] \leq 1 \\
& r[n] = r_a[n] + r_s[n], \quad n = 1, \dots, N_I \\
& r[n] \leq \text{PPFD}_{\text{max}}, \quad n = 1, \dots, N_I \\
& r[n] - \text{PPFD}_{\text{min}} \; v[n] \geq 0, \quad n = 1, \dots, N_I \\
& \sum_{n=1}^{N_I} 36 \times 10^{-4} \; r[n] \;  = \text{DLI}_{\text{Curtl}} \\
& u_{\text{opt}}[i] = u[i], \quad x_{\text{opt}}[i] = x[i], \quad r_{a_{\text{opt}}}[n]
\end{flalign}
$Y[n]$ combines real values for $n < i$ and predicted values for $n \geq i$ for solar radiation. $r[n]$ represents the lighting recipe, consisting of two components: $r_s[n]$ (solar radiation) and $r_a[n]$ (artificial lighting). $r_s[n]$ includes a binary variable $v[n]$, indicating the presence or absence of sunlight, which depends on weather and cannot be controlled. $w[n]$ adjusts solar input through shading. Two key constraints apply: maintaining $r[n]$ within the PPFD range and ensuring that the 24-hour sum of $r[n]$ meets $\text{DLI}_{\text{Curtl}}$. $r_{a_{\text{opt}}}[n]$ denotes the optimized artificial lighting component for online control, composed of $u_{\text{opt}}[i]$ and $x_{\text{opt}}[i]$. While $I_{D_{max}}$ and $I_{L_{max}}$ may be relevant for flowering species, they are typically not included as constraints in the objective function.

\begin{figure}
	\centering
	\scalebox{0.13}{\includegraphics{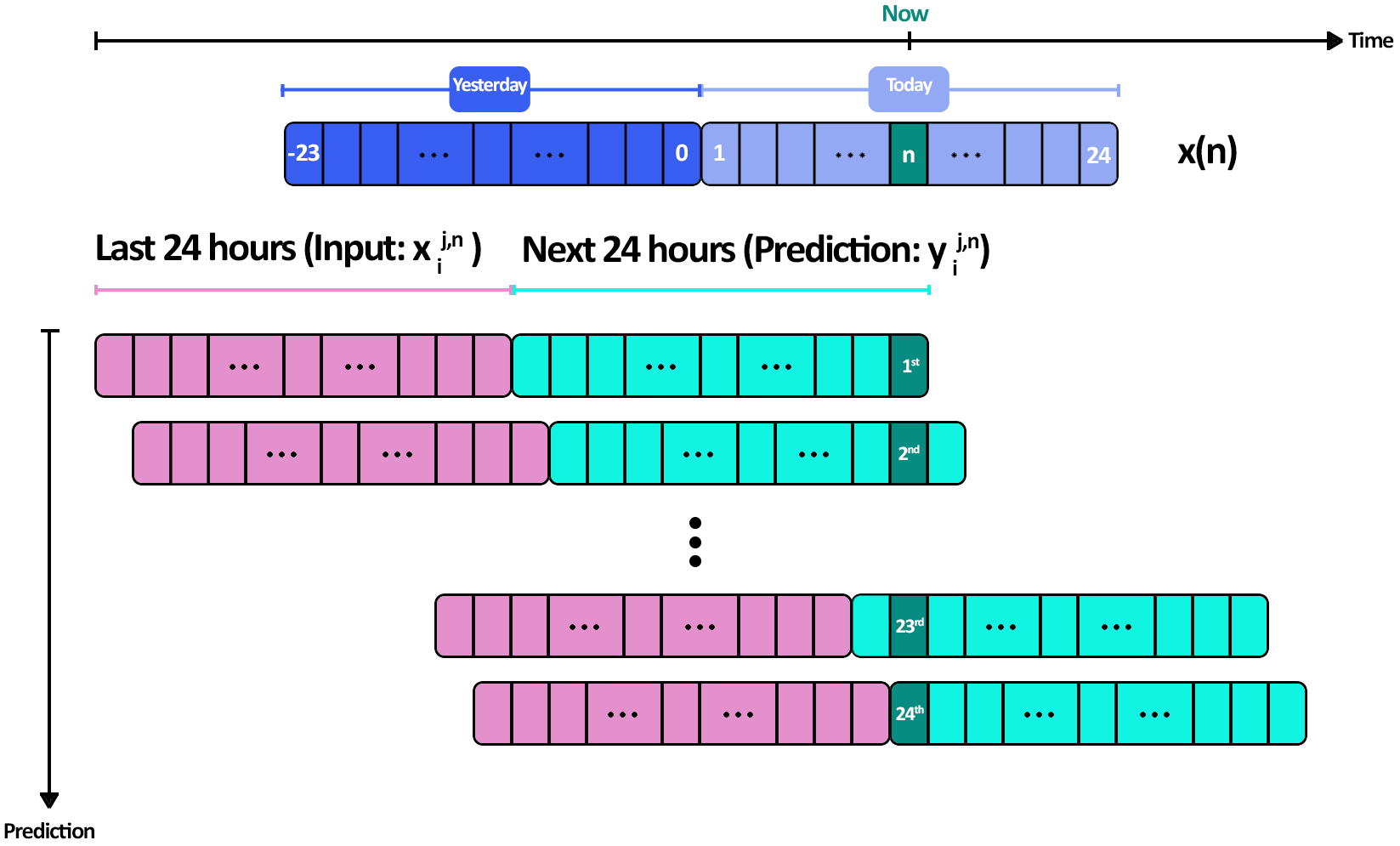}}
	\caption{Overall procedure for online prediction of electricity price and solar radiation.}
	\label{fig4}
\end{figure}

As mentioned, we use two separate transformer-based predictors (Section \ref{Predicions}) to forecast the next 24 hours of solar radiation and electricity prices using the given features. For each hour, we use all available predictions. As shown in Fig. \ref{fig4}, there are 24 predictions for each hour. Since older predictions rely on older data, we apply a weighting function (Equation (4)) to give more importance to recent ones.
\begin{flalign}
w_n = \frac{e^{\alpha n}}{\sum_{j=0}^{N_I}e^{\alpha j}}, \quad n = 1, \dots, N_I 
\end{flalign}
where $w_n$ is the weight assigned to index $n$, and $\alpha$ is a scaling factor that controls the steepness of the exponential growth. 
\begin{figure}[t]
	\centering
	\scalebox{0.42}{\includegraphics{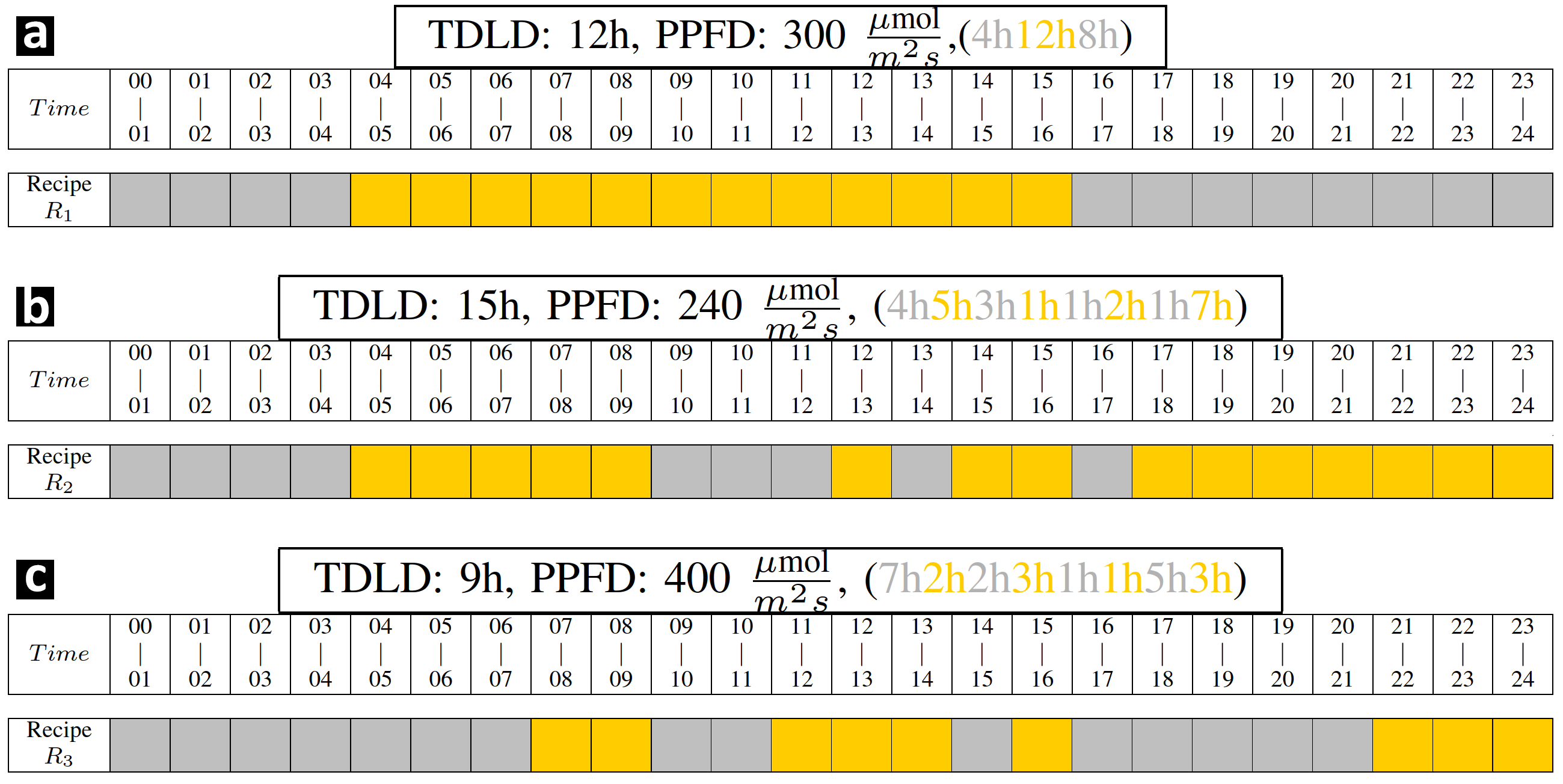}}
	\caption{Trial recipes for plant growth experiments. Yellow cells are light intervals while grey cells are dark intervals.}
	\label{fig5}
\end{figure}
\begin{figure}[t]
	\centering
	\scalebox{0.07}{\includegraphics{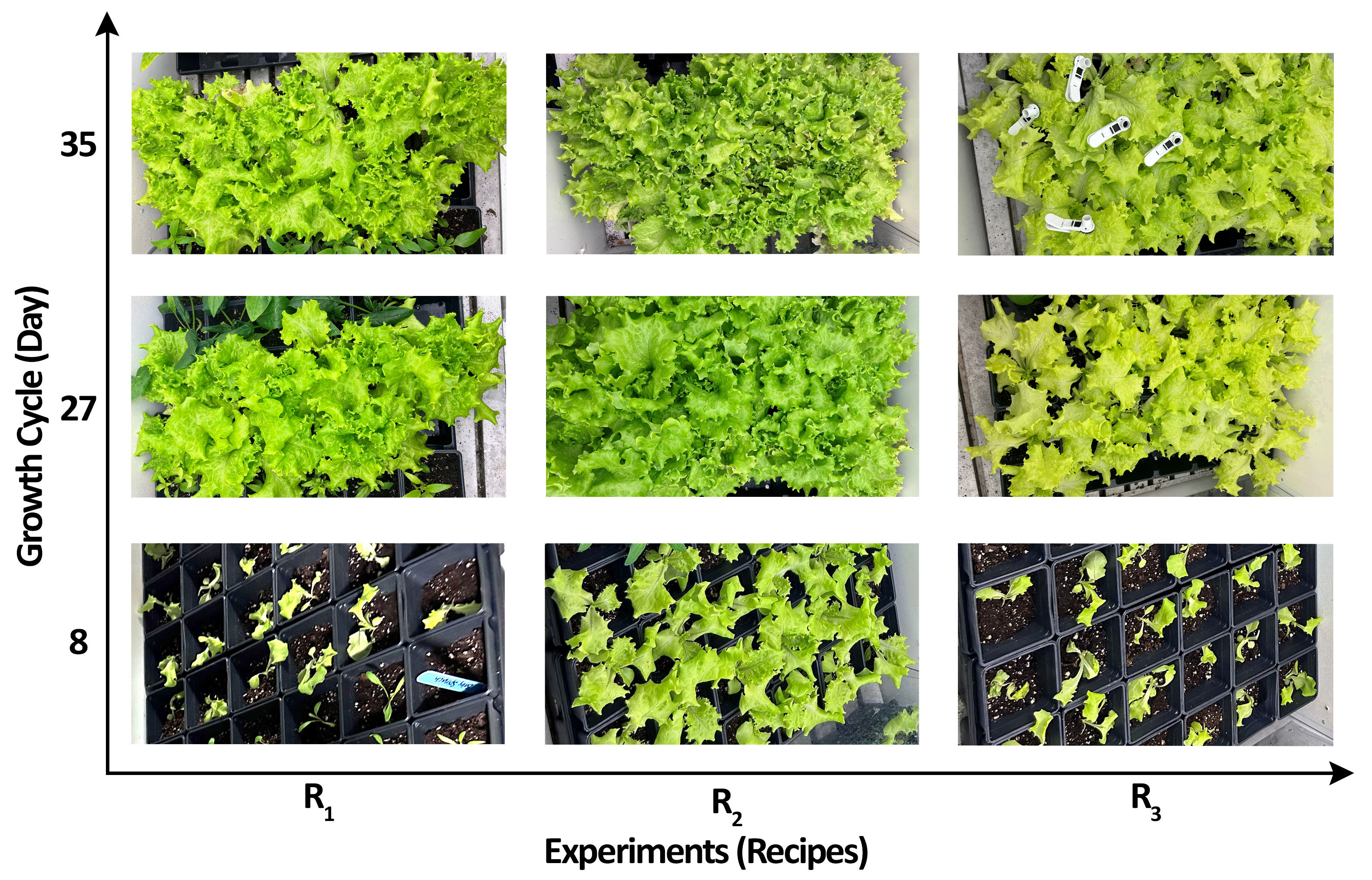}}
	\caption{Results of trial recipes across different growth stages.}
	\label{fig6}
\end{figure}
\section{Results and Discussion}
This section presents the results from the real-world experiments of the designed trial recipes, as well as simulation results on a one hectare greenhouse.
\subsection{Experimental Results for Plant Boundary Identification}
In this subsection, we outline the experimental setup and key results for the trial recipes and discovery of boundary conditions for minimum DLI. Experiments were conducted using plant growth chambers (model E8, Conviron) with a 46-inch growth height and an 8-square-foot cultivation area, as well as a combination of T8 fluorescent lamps, delivering up to 575 $\frac{\mu \text{mol}}{m^2 \cdot s}$ with uniform distribution via a counterbalanced light canopy. The crops under test were loose-leaf lettuce (\textit{Lactuca sativa}). Based on the literature \cite{kang2013light}, the selected parameters included a DLI of 12–19 $\frac{\text{mol}}{m^2 \cdot \text{day}}$, TDLD of 6-24 hours, and PPFD of 130–880 $\frac{\mu \text{mol}}{m^2 \cdot s}$. Light and dark intervals ranged from 1–24 hours and 1–18 hours, respectively. To approach the lower bound of the DLI range and assess the impact of different light/dark distributions on plant health, we maintained a constant DLI of 12.96 $\frac{\text{mol}}{m^2 \cdot \text{day}}$ across three chambers. As shown in Fig. \ref{fig5}, three lighting recipes were applied: a control (R1), a lower PPFD (R2), and a higher PPFD (R3). TDLDs were set to 12, 15, and 9 hours, respectively, with PPFD values adjusted accordingly to maintain the fixed DLI. The ON and OFF interval durations were selected to cover a wide range of OFF intervals (1 to 7 hours) within $R_2$ and $R_3$. To evaluate the potential impact of the designed recipes on plant health, quantitative and qualitative measurements were taken, including leaf size measurement by a ruler, $Fv/Fm$ measurements by a portable chlorophyll fluorometer (model OS30p+, Opti-Sciences), and leaf color by a digital camera.

\begin{figure}[t]
        \centering
	\scalebox{0.36}{\includegraphics{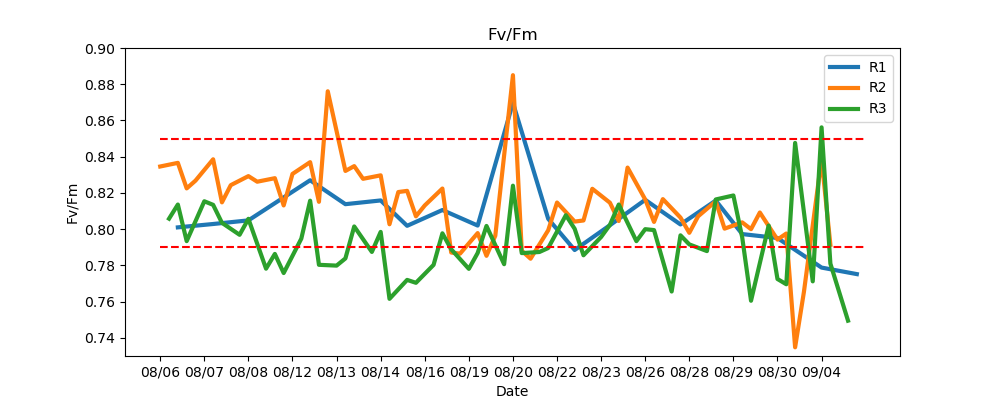}}
	\caption{Time-domain plot of $F_v/F_m$ across different lighting recipes.}
	\label{fig7}
\end{figure}

As seen in Table \ref{table1}, leaf area measurements were taken for three samples per chamber at two time points (days 20 and 30 of the growth cycle). The results indicate that lettuce grown under $R_2$ had the largest leaf area, with approximately 9\% more growth than $R_1$ in the final stage, while $R_3$ showed a 54\% reduction compared to $R_2$. These findings confirm that a longer photoperiod with lower PPFD supports greater leaf expansion. Additionally, based on Fig. \ref{fig6}, leaf color assessments indicate that plants grown under $R_1$ and $R_2$ appeared fresher and more harvest-ready than those under $R_3$. These findings underscore the benefits of a longer photoperiod in enhancing lettuce growth and quality. Lastly, $F_v/F_m$ measurements were recorded for all three recipes as seen in Fig. \ref{fig7}, which illustrates the time-domain evolution of $F_v/F_m$ across the growth cycle, with stress levels particularly low during later stages. Most $F_v/F_m$ values remained within the healthy range (0.79–0.85), indicating minimal stress and confirming the effectiveness of the light curtailment strategies. Notably, $R_2$ exhibited the most favorable stress profile, consistent with its superior plant quality.
\begin{table}[t]
	\centering
	\caption{Numerical assessment of the experiment.}
	\begin{adjustbox}{width=0.3\textwidth}
		\begin{tabular}{|c|c|c|c|}
			\hline
    		\multirow{2}{*}{Day} & \multicolumn{3}{c|}{Average Leaf Area ($cm^2$)} \\ 
                \cline{2-4}
                & $R_1$ & $R_2$ & $R_3$ \\
    		\hline
                \hline
    		$20^{th}$ & 52.01 & 82.75 & 26.73 \\
    		\hline
    		$30^{th}$ & 98.83 & 108.81 & 50.45 \\
    		\hline
		\end{tabular}
	\end{adjustbox}
	\label{table1}
\end{table}
These findings suggest that even with reduced DLI and frequent light interruptions, plants maintained healthy photosynthetic function. This supports the feasibility of integrating such boundary conditions into the optimal design of daily lighting recipes.

\subsection{Simulation Results}
We evaluate the efficacy of our proposed system by simulating a greenhouse model as described in Section II. Simulation parameters include an area of 10,000 $m^2$ growing loose-leaf lettuce crops (\textit{Lactuca sativa}), with a  DLI of 12.96 $\frac{\text{mol}}{m^2 \cdot \text{day}}$. The internal climate was maintained with daytime and nighttime temperature ranges of [20, 24]°C and [12, 16]°C, respectively, and relative humidity between 60–70\%. The average $\text{CO}_2$ concentration was regulated at 820 ppm, while cover transmittance was set to 0.6. The rated power for individual loads, including fans, dehumidification, fogging systems, LEDs, $\text{CO}_2$ injectors, chillers, and heaters, was 0.13, 2.2, 2.2, 0.6, 8, 6.6, and 3.3 kW, respectively.  The training dataset, gathered from 2015 to 2023 for Ontario \cite{ECCC}, was used to develop the proposed system, which was tested using 2024 data \cite{ECCC}. Electricity prices are obtained from the Class B structure within Ontario system \cite{IESO}. We compare our proposed system with a baseline recipe from literature \cite{kang2013light}, with a DLI of 15.03 $\frac{\text{mol}}{{m^2} \cdot \text{day}}$, which corresponds to a 12-hour lighting recipe with a PPFD of 348 $\frac{\mu \text{mol}}{{m^2}s}$ from 6:00-18:00, supplemented by solar radiation.



\begin{table*}[ht]
	\centering
	\caption{Summary of Energy Consumption, Costs, and Charges for 2024}
	\label{tab:summary}
	\resizebox{\textwidth}{!}{%
		\begin{tabular}{|c|cc|cc|cc|cc|cc|cc|cc|}
			\hline
			\multirow{2}{*}{\centering Month} & \multicolumn{2}{c|}{Energy Consumption (MWh)} & \multicolumn{2}{c|}{Peak (MW)} & \multicolumn{2}{c|}{Energy Price (K CAD)} & \multicolumn{2}{c|}{Peak Demand Charge (K CAD)} & \multicolumn{2}{c|}{GA Cost (K CAD)} & \multicolumn{2}{c|}{Total Cost (K CAD)} & \multicolumn{2}{c|}{Energy and Cost Reduction} \\
			\cline{2-15}
			& Baseline & Optimized & Baseline & Optimized  & Baseline & Optimized & Baseline & Optimized & Baseline & Optimized & Baseline & Optimized & Energy (\%)& Cost (\%)\\
			\hline
			Jan & 1,253.1 & 1,096.6 & 5.6  & 4.0  & 57.5 & 47.3 & 60.1 & 42.9 & 62.8 & 54.9 & 180.4	& 145.1 & 12.49	& 19.57 \\
			Feb & 852.7 & 689.6 & 4.6 & 3.4 &  25.9 & 19.4 & 49.9 & 36.9 & 72.5 & 58.6 & 148.3 & 114.9 & 19.13 & 22.52 \\
			Mar & 731.6 & 569.7 & 4.3 & 3.6 &  21.9 & 15.9 & 47.0 & 39.1 & 51.9 & 40.4 & 120.8 & 95.4 & 22.13 & 21.03 \\
			Apr & 594.4 & 444.9 & 3.7 & 3.0 &  18.0 & 11.9 & 40.3 & 32.6 & 39.4 & 29.5 & 97.7	&74.0	&25.15	&24.26 \\
			May & 603.1 & 442.3 & 3.9 & 3.5 &  18.6 & 13.7 & 42.4 & 37.9 & 27.7 & 20.3 & 88.7	&71.9	&26.66	&18.94  \\
			Jun & 596.5 & 469.6 & 4.1 & 3.8 &  19.8 & 16.3 & 44.2 & 40.9 & 39.6 & 31.1 & 103.6	&88.3	&21.27	&14.77  \\
			Jul & 679.9 & 540.2 & 3.4 & 2.8 &  26.3  & 22.5 & 36.5 & 30.4 & 55.6 & 44.1 & 118.4	&97.0	&20.55	&18.07 \\
			Aug & 656.5 & 505.8 & 4.0 & 3.5 &  25.0  & 20.8 & 43.3 & 37.7 & 48.8 & 37.6 & 117.1	&96.1	&22.96	&17.93 \\
			Sep & 588.6 & 419.9 & 3.6 & 3.0 &  19.6  & 13.9 & 39.3 & 33.0 & 45.7 & 32.6 & 104.6	&79.5	&28.66	&24.00 \\
			Oct & 663.9 & 487.8 & 4.0 & 3.1 &  25.8  & 15.9 & 43.7 & 33.8  & 52.0 & 38.2 & 121.5	&87.9	&26.53	&27.65 \\
			Nov & 925.1 & 752.5 & 4.4 & 3.5 &  30.5 & 20.5 & 47.4 & 38.2 & 58.9 & 47.9 & 136.8	&106.6	&18.66	&22.08\\
			Dec & 1,228.9 & 1,065.0 & 5.2 & 3.6 & 50.5 & 40.2 & 55.8 & 39.3 & 77.7 & 67.3 & 184.0	&146.8	&13.34	&20.22 \\
			\hline
			Annual & 9,374.3 & 7,483.9 & 5.6 & 4.0 & 339.4 (22.3\%) & 258.3 (21.5\%) & 549.9 (36.1\%) & 442.7 (36.8\%) & 632.6 (41.6\%) & 502.5(41.7\%) & 1,521.9 & 1,203.5 & \textbf{20.17} & \textbf{20.92} \\
			\hline
		\end{tabular}%
	}
\end{table*}

\subsubsection{Case Study I: Electricity Price and Solar Radiation Predictions} 
\begin{figure}[t]
	\scalebox{0.17}{\includegraphics{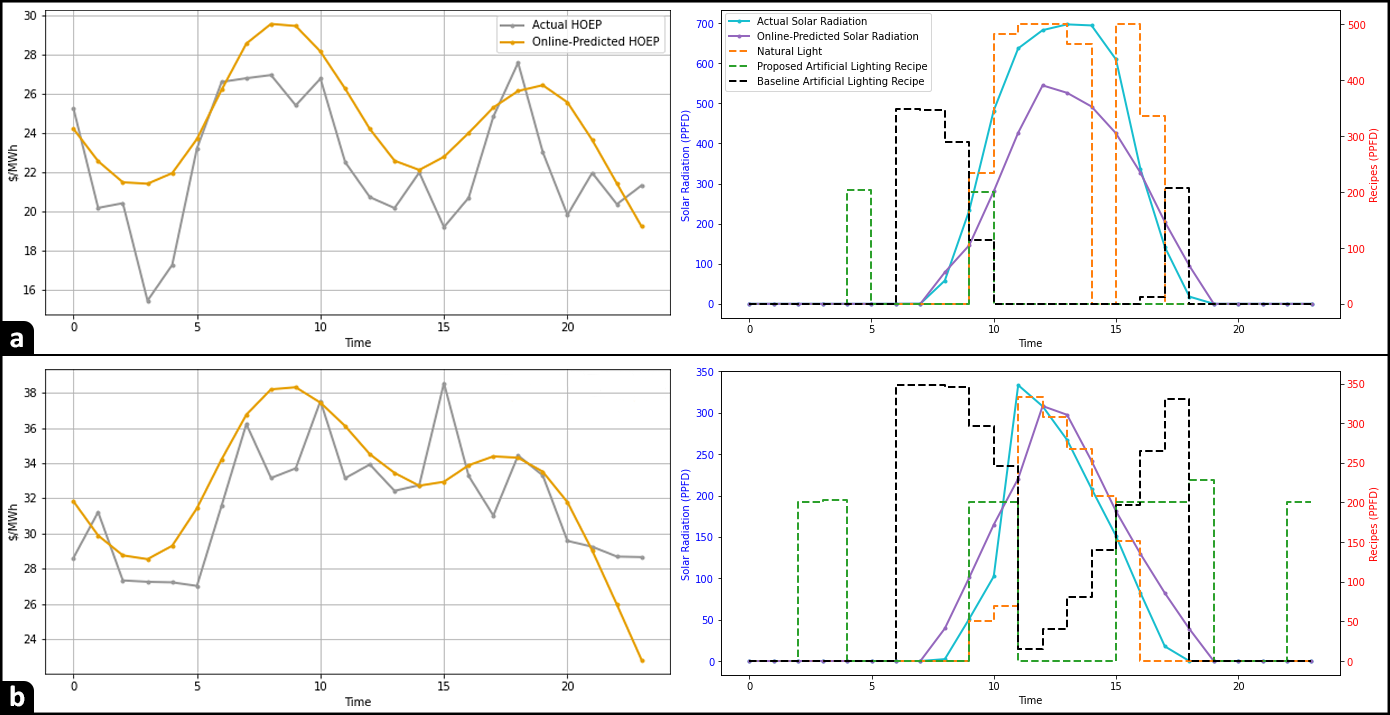}}
	\caption{Results of predictions  and optimization of lighting recipes.}
	\label{fig8}
\end{figure}

This case study analyzes the impact of the predictors on the generated recipe. 
Fig. \ref{fig8} presents two pairs of plots (a, b), with each pair containing two subplots. The left subplot illustrates electricity price prediction performance by comparing the actual hourly Ontario electricity price (HOEP) with the online-predicted HOEP. The right subplot features two axes, displaying solar radiation and lighting recipes, both in PPFD. Solid lines represent actual and online-predicted solar radiation values, while dashed lines depict the natural light, proposed artificial lighting recipe, and the baseline artificial lighting recipe.


Fig \ref{fig8}-a shows a day where solar radiation intensity is high and the sun provides a significant portion of the required DLI. The accurate prediction of solar radiation thus enables the optimized approach to reduce artificial lighting to other periods in contrast to the baseline artificial lighting recipe. Additionally, the accurate electricity price prediction has facilitated the distribution of proposed artificial lighting intensity during low-tariff periods (4:00-5:00) while also minimizing peak demand.


Fig. \ref{fig8}-b illustrates a similar situation where solar radiation is low, a fact that is gradually identified through solar radiation prediction. Consequently, more artificial lighting is required in both the baseline and optimized recipes. The proposed method optimally distributes artificial lighting to maximize the use of available solar radiation while scheduling artificial lighting during low-tariff periods (2:00-5:00) based on price prediction.

\begin{figure}[t]
	\scalebox{0.34}{\includegraphics{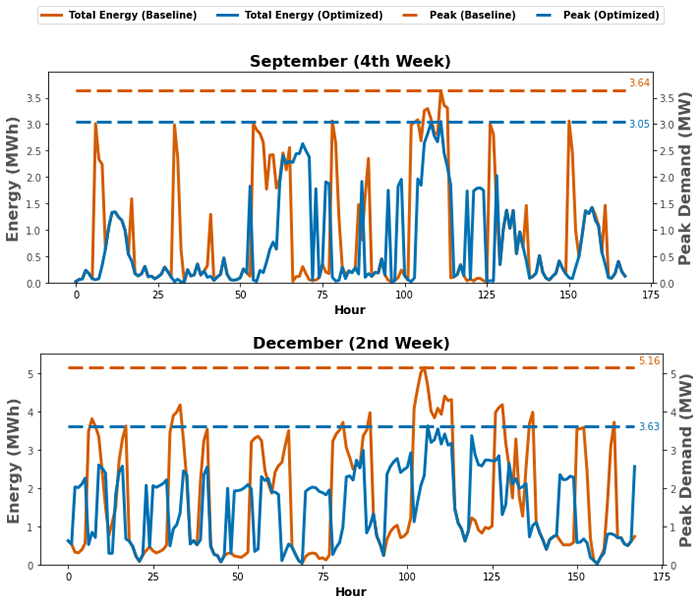}}
	\caption{Two week energy profile for baseline vs optimized system.}
	\label{fig9}
\end{figure}
The results demonstrate that the proposed method effectively schedules artificial lighting during low-tariff periods. In contrast, the baseline artificial lighting recipe lacks flexibility in PPFD, photoperiod, and light/dark cycles, and is unable to fully utilize available solar radiation due to a fixed PPFD cap of 348. Consequently, the optimized strategy achieves a 45.08\% reduction in energy costs (based solely on electricity prices) and a 60.13\% decrease in artificial light usage over the course of 2024.
\subsection{Case Study II: Monthly Energy Profile Analysis}
This subsection presents a comparative analysis of the greenhouse’s energy profile, highlighting differences between the baseline scenario and the optimized lighting recipe. The hourly energy profiles for two sample weeks (September and December) are shown in Fig. \ref{fig9}, while monthly results for energy costs, peak load, and total energy cost reductions are shown in Table \ref{tab:summary}. Fig. \ref{fig9} illustrates a significant reduction in both energy consumption and peak demand for a shoulder/transitional month (September), as well as a winter month (December). Table \ref{tab:summary} further reaffirms these facts, with reductions in energy consumption, peak load, and GA in all 12 months of the year. Annually, this translates to an 1890.4 MWh reduction in energy consumption (20.17\%), \$318.4K reduction in energy costs (20.92\%), and a peak load reduction of 1.4 MW (33.3\%).

The impact of the algorithm varies by season. During the warmer months (May to August), the cooling system contributes peaks comparable to those of the lighting system, and their combined demand approaches the baseline peak. This is because solar radiation supplies a substantial portion of the required light, reducing the effect of the lighting strategy. However, as shown in Table \ref{tab:summary}, energy and cost reductions remain noticeable, ranging from 20.55\%–26.66\% and 14.77\%–18.94\%, respectively. Additionally, peak demand reduction ranges between 7.31\% and 17.64\%. In winter months with colder temperatures and less solar radiation, the heating demand as well as the reliance on artificial lighting increases. In these months, the effect of the proposed lighting strategy on peak demand becomes more pronounced, as seen in Fig. \ref{fig9} when comparing the September plot with the December plot.  Lastly, in transitional months such as April, September, and October, the energy and cost reductions are higher than in other months, mostly due to the reduced need for excessive heating/cooling, enabling the proposed lighting system to find more opportunities for cost savings. 


\section{Conclusion}
We proposed a comprehensive lighting control strategy aimed at enhancing the energy efficiency and economic performance of indoor lettuce cultivation. By integrating light curtailment with advanced control of light intensity and photoperiod, the approach leverages the physiological tolerance of plants to brief light interruptions. Two transformer-based neural networks were developed to accurately forecast 24-hour ahead solar radiation and electricity prices, enabling a model predictive control framework to generate optimized, cost-aware lighting recipes that account for available solar input. The strategy effectively shifts high-energy lighting operations to low-tariff periods while strictly adhering to plant physiological constraints, including DLI and appropriate PPFD ranges. Simulation results for a one-hectare greenhouse, based on real-time electricity pricing data from the Ontario market, demonstrated an annual cost reduction of \$318,400 (20.92\%), an average peak load reduction of 833 kW (19.62\%), and total energy savings of 20.17\%. These findings highlight the potential of intelligent lighting management to deliver both economic and environmental benefits in modern indoor farming systems. Future work includes the expansion of the proposed framework into contracted grid services for peak load reduction, as well as integrating other DERs to further enhance its efficacy.

\bibliographystyle{IEEEtran}

\bibliography{citations}

\end{document}